\documentclass{ws-ijmpd}
\usepackage[super,compress]{cite}
\usepackage{color}
\newcommand{\eq}[1]{(\ref{#1})}
\begin{document}

\markboth{Hyun Seok Yang}
{Emergent Spacetime for Quantum Gravity}

%
\catchline{}{}{}{}{}
%

\title{EMERGENT SPACETIME FOR QUANTUM GRAVITY}

\author{HYUN SEOK YANG}

\address{Center for Quantum Spacetime, Sogang University, Seoul
121-741, Korea \\
hsyang@sogang.ac.kr}

\maketitle

\begin{history}
\received{31 January 2016}
\end{history}

\begin{abstract}
We emphasize that noncommutative (NC) spacetime necessarily implies emergent spacetime
if spacetime at microscopic scales should be viewed as NC.
In order to understand NC spacetime correctly, we need to deactivate the thought patterns that
we have installed in our brains and taken for granted for so many years.
Emergent spacetime allows a background-independent formulation of quantum gravity that
will open a new perspective to resolve the notorious problems in theoretical physics
such as the cosmological constant problem, hierarchy problem, dark energy, dark matter,
and cosmic inflation.
\end{abstract}

\keywords{Emergent spacetime; Quantum gravity; Noncommutative spacetime}

\ccode{PACS numbers: 04.60.-m; 11.25.Tq; 02.40.Gh}


\section{What Every Physicist Should Know about Noncommutative Spacetime}

Since Newton the concept of space and time has gone through various changes. In retrospect,
the upheaval of new physics has accompanied the novel concept of space and time.
And history is a mirror to the future. The concept of emergent gravity and spacetime
recently activated by the AdS/CFT correspondence \cite{ads-cft1,ads-cft2,ads-cft3}
advocates that spacetime is not a fundamental entity existed from the beginning
but an emergent property
from a primal monad such as matrices. However physicists have still devoted themselves
to the 20th century physics and have been complacent in the contemporary paradigm
although the concept of emergent spacetime asks us to go beyond the horizon of
the contemporary physics. It is our belief that the beginning of the 21st century is
the right time to go beyond the spacetime concept developed for the 20th century physics
and the emergent spacetime is a significant new paradigm for the 21st century physics
which may shed light on the origin of spacetime and our Universe.

In string theory, two exclusive spacetime pictures exist together although they are conceptually
in deep discord with each other. One is founded on the general relativity while
the other is based on the emergent spacetime. However these two spacetime pictures
are exclusive and irreconcilable each other. Thus it is necessary to formulate string theory
entirely based on a unique spacetime picture. Since we think that the emergent spacetime
picture is a more fundamental approach, we want to understand string theory
based on the novel spacetime picture. Indeed string theory basically implies the emergent
spacetime picture. (See the quotation in Sec. IV in Ref. \refcite{hsy-review}.)
However the current string theory assumes from the outset that strings are vibrating
in a preexisting spacetime and so resides in the territory of general relativity.
After all, a spacetime is inserted into the theory ``by hand" and thus
spacetime does not emerge from a fundamental ingredient in the theory.
Hence we need a new approach to string theory
in order to realize the concept of emergent spacetime.

String theory is defined by replacing particles (point-like objects) with strings
(one-dimensional objects). In order to do this, we need to introduce a \textit{new}
constant $\alpha'$ whose physical dimension is $(\mathrm{length})^2$.
It is well-known that the new constant $\alpha'$ introduces a new duality depicted
by $R \to R' = \frac{\alpha'}{R}$. This is known as the T-duality in string theory,
but it is impossible in particle theories $(\alpha'=0)$.
It is important to notice that a new physical constant such as $\hbar$ and $\alpha'$
introduces a deformation of some structure in a physical theory \cite{hsy-jhep09,hsy-review}.
For instance, the Planck constant $\hbar$ in quantum mechanics carries the physical dimension
$ [\hbar] = (\mathrm{length}) \times (\mathrm{momentum})$ and so it deforms the algebraic
structure of particle phase space from commutative to noncommutative (NC) space, i.e.,
\begin{equation}\label{nc-particle}
    x p - p x = 0 \qquad \Rightarrow \qquad x p - p x = i \hbar.
\end{equation}
An educated reasoning motivated by the fact that $[\alpha'] = (\mathrm{length})
\times (\mathrm{length})$ leads to a natural speculation that $\alpha'$ brings
about the deformation of the algebraic structure of spacetime itself such that
\begin{equation}\label{nc-space}
    x y - y x = 0 \qquad \Rightarrow \qquad x y - y x = i \alpha'.
\end{equation}
From the deformation theory point of view, replacing particles with strings
is equivalent to the transition from commutative space to NC space.
This may be supported by the fact that the NC space \eq{nc-space} defines only a minimal area
whereas the concept of point is doomed as we have learned from quantum mechanics defined
by the NC phase space \eq{nc-particle}. The minimal surface in the NC space \eq{nc-space}
acts as a basic building block of string theory. Remarkably the deformation \eq{nc-space} provides
us an important clue for a background-independent formulation of string theory \cite{part1}.

Since the mathematical structure of the NC space \eq{nc-space} is essentially the same as
quantum mechanics in Eq. \eq{nc-particle}, we will consider the deformation \eq{nc-space}
to be quantization. An important question is whether the quantization \eq{nc-space} in terms
of $\alpha'$ can be regarded as an alternative formulation of string theory so that
it realizes all essential aspects of string theory.
The recent results \cite{hsy-jhep09,hsy-review,part1,q-emg} suggest that the answer is yes
and moreover it naturally realizes the concept of
emergent spacetime which was elusive so far in the usual string theory.
In this approach quantum gravity is formulated as a dynamical theory of NC spacetime
and the NC spacetime necessarily implies emergent spacetime \cite{q-emg}.
Therefore the NC space \eq{nc-space} denoted by $\mathbb{R}^{2}_{\alpha'}$
is much more radical and mysterious than we thought.
In order to understand NC spacetime correctly, we need to deactivate the thought patterns
that we have installed in our brains and taken for granted for so many years.

The reason is the following.
Recall that the NC phase space \eq{nc-particle} in quantum mechanics introduces
the wave-particle duality. Indeed the NC space \eq{nc-space} also brings about
a radical change of physics since the NC nature of spacetime is responsible
for a new kind of duality, known as the gauge-gravity duality.
A primary cause of the radical change of physics in quantum mechanics
is that the NC phase space \eq{nc-particle} introduces a {\it complex} vector
space called the Hilbert space. This is also true for
the NC space \eq{nc-space} since its mathematical structure is
essentially the same as quantum mechanics. Moreover the NC space $\mathbb{R}^{2}_{\alpha'}$
admits a nontrivial inner automorphism.
For example, for an arbitrary NC field $f(x, y)$, we have the relation given by
\begin{equation}\label{tr-auto}
    f(x+a, y) = U(a)^\dagger f(x, y) U(a), \qquad f(x, y + b) = U(b)^\dagger f(x, y) U(b)
\end{equation}
where $U(a) = \exp(- \frac{ i a y}{\alpha'})$ and $U(b) = \exp(\frac{ i b x}{\alpha'})$.
A striking feature of the NC space is thus that every points are unitarily equivalent
because translations in $\mathbb{R}^{2}_{\alpha'}$ are simply a unitary transformation
acting on the Hilbert space $\mathcal{H}$.
As a result, the concept of space is doomed and the classical space is replaced by a state in
the Hilbert space $\mathcal{H}$. This fact leads to an important picture that
classical spacetime is an emergent concept derived from a NC algebra.
In other words, NC spacetime necessarily implies emergent spacetime
if spacetime at microscopic scales should be viewed as NC.
Moreover any dynamical variable defined on $\mathbb{R}^{2}_{\alpha'}$
becomes an operator acting on the Hilbert space $\mathcal{H}$. In particular, any NC field
can be regarded as a linear operator acting on the Hilbert space.
Note that the NC space \eq{nc-space} is equivalent to the Heisenberg algebra of harmonic
oscillator, i.e. $[a, a^\dagger] = 1$, if the annihilation operator is defined by
$a = \frac{1}{\sqrt{2 \alpha'}} (x+iy)$.
Thus the Hilbert space for $\mathbb{R}^{2}_{\alpha'}$ is the Fock space and has a countable
basis. Then the representation of NC fields on the Hilbert space $\mathcal{H}$ is
given by $N \times N$ matrices where $N = \mathrm{dim}(\mathcal{H}) \to \infty$.
Consequently, the NC space \eq{nc-space} leads to an interesting equivalence between a
lower-dimensional large $N$ gauge theory and a higher-dimensional NC $U(1)$ gauge theory \cite{q-emg}.

\begin{figure}
\centerline{\psfig{file=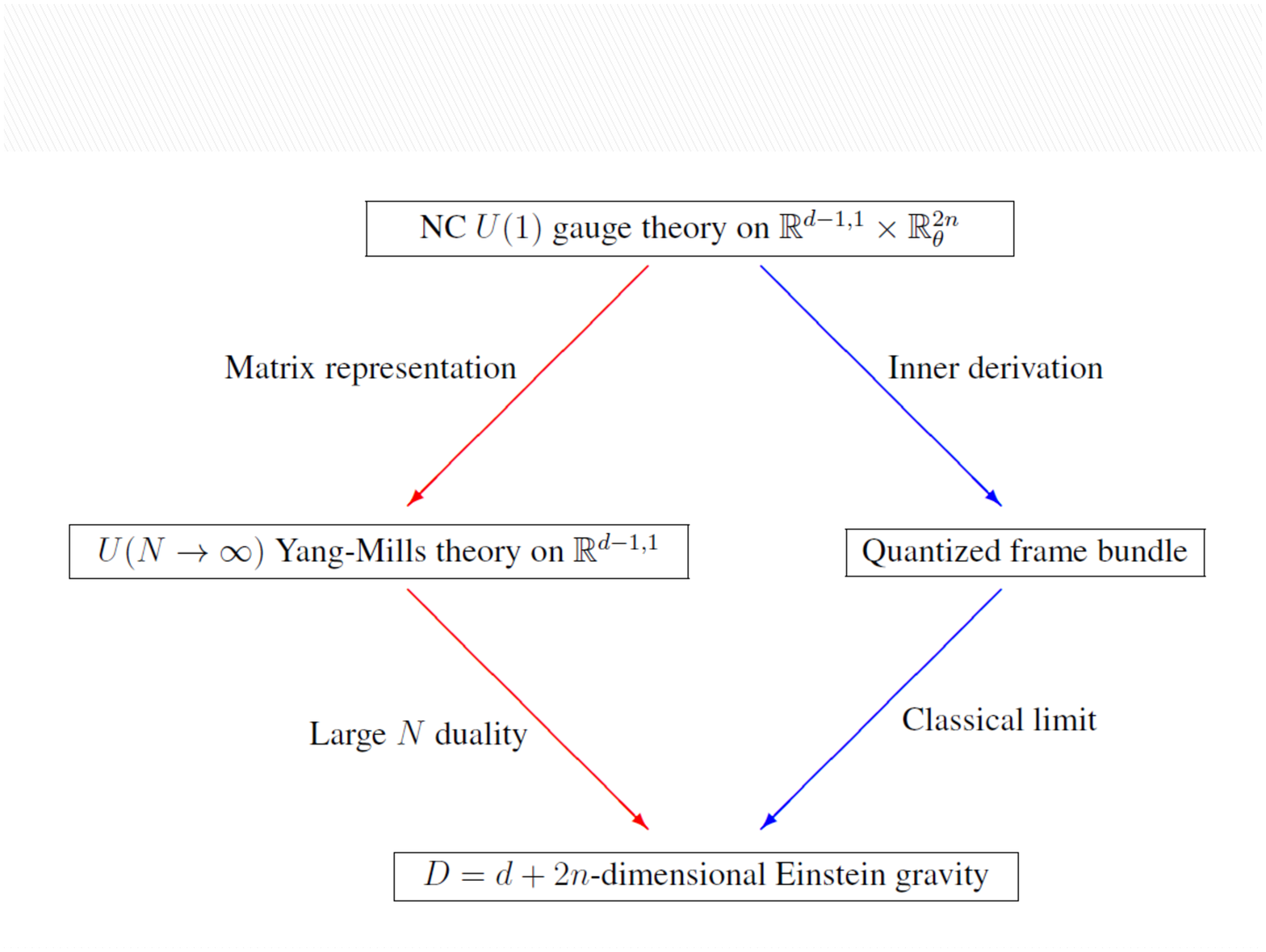,width=12.7cm}}
\vspace*{18pt}
\caption{Flowchart for emergent gravity \cite{part1}}
\label{fchart:emg}
\end{figure}

In next section we will show that $U(1)$ gauge theory defined on a NC spacetime such as \eq{nc-space}
defines quantum gravity according to the large $N$ duality as depicted in Fig. \ref{fchart:emg}.
We thus demonstrate how the deformation \eq{nc-space} provides
us a background-independent formulation of string theory \cite{part1}.

\section{Emergent Spacetime from Large $N$ Duality}

Let us consider a $2n$-dimensional NC space denoted by $\mathbb{R}^{2n}_{\theta}$
whose coordinate generators obey the commutation relation
\begin{equation}\label{extra-nc2n}
    [y^a, y^b] = i \theta^{ab}, \qquad a, b = 1, \cdots, 2n,
\end{equation}
where $(\theta)^{ab} = \alpha' (\mathbf{1}_n \otimes i \sigma^2)$ is a $2n \times 2n$ constant
symplectic matrix and $l_s \equiv \sqrt{\alpha'}$ is a typical length scale set by the vacuum.
The NC space $\mathbb{R}^{2}_{\alpha'}$ corresponds to the $n=1$ case.
Similarly to the $n=1$ case, the NC space (\ref{extra-nc2n}) is equivalent to the Heisenberg algebra
of $n$-dimensional harmonic oscillator. Hence the underlying Hilbert space on which the NC
$\star$-algebra $\mathcal{A}_\theta$ acts is given by the Fock space defined by
\begin{equation}\label{fock-space}
    \mathcal{H} = \{|\vec{n}\rangle \equiv |n_1, \cdots, n_n \rangle | \; n_i \in \mathbb{Z}_{\geq 0},
    \; i=1, \cdots, n \},
\end{equation}
which is orthonormal, i.e., $\langle \vec{n}|\vec{m} \rangle = \delta_{\vec{n},\vec{m}}$ and
complete, i.e., $\sum_{\vec{n} = 0}^{\infty} | \vec{n}\rangle \langle \vec{n}| = \mathbf{1}_{\mathcal{H}}$,
as is well-known from quantum mechanics. Since the Fock space (\ref{fock-space}) has a countable basis,
it is convenient to introduce a one-dimensional basis using the ``Cantor diagonal method" to put the
$n$-dimensional non-negative integer lattice in $\mathcal{H}$ into one-to-one correspondence
with the natural numbers:
\begin{equation}\label{cantor}
\mathbb{Z}^n_{\geq 0} \leftrightarrow \mathbb{N}: |\vec{n}\rangle \leftrightarrow |n \rangle, \;
n = 1, \cdots, N \to \infty.
\end{equation}
In this one-dimensional basis, the completeness relation of the Fock space (\ref{fock-space}) is now
given by $\sum_{n = 1}^{\infty} | n \rangle \langle n| = \mathbf{1}_{\mathcal{H}}$.

The Hilbert space for $\mathbb{R}^{2n}_{\theta}$ is also the Fock space and so
the representation of NC fields on the Hilbert space $\mathcal{H}$ is
given by $N \times N$ matrices where $N = \mathrm{dim}(\mathcal{H}) \to \infty$.
Consider two arbitrary dynamical fields $f(x,y)$ and $g(x,y)$ on a $D=(d+2n)$-dimensional
NC spacetime $\mathbb{R}^{d-1,1} \times \mathbb{R}^{2n}_{\theta}$
which are elements of the NC $\star$-algebra $\mathcal{A}_\theta^d \equiv \mathcal{A}_\theta
\big( C^\infty (\mathbb{R}^{d-1,1}) \big) = C^\infty (\mathbb{R}^{d-1,1}) \otimes
\mathcal{A}_\theta$. Since the dynamical variables in $\mathcal{A}_\theta^d$ can be regarded
as operators acting on the Hilbert space (\ref{fock-space}), we can represent the operators
in the Fock space (\ref{fock-space}) as $N \times N$ matrices in $\mathrm{End}(\mathcal{H})
= \mathcal{A}_N^d \equiv \mathcal{A}_N \big( C^\infty (\mathbb{R}^{d-1,1}) \big)
= C^\infty (\mathbb{R}^{d-1,1}) \otimes \mathcal{A}_N$ where $N = \mathrm{dim}(\mathcal{H}) \to \infty$:
\begin{eqnarray}	
\begin{array}{rcl}
     && f(x,y) = \sum_{n,m=1}^\infty | n \rangle \langle n| f (x,y) | m \rangle \langle m|
      := \sum_{n,m=1}^\infty F_{nm} (x) | n \rangle \langle m|, \\
     && g(x,y) = \sum_{n,m=1}^\infty | n \rangle \langle n| g (x,y) | m \rangle \langle m|
      := \sum_{n,m=1}^\infty G_{nm} (x) | n \rangle \langle m|,
\end{array}
\label{matrix-rep}
\end{eqnarray}
where $F(x)$ and $G(x)$ are $N \times N$ matrices in $\mathcal{A}^d_N = \mathrm{End}(\mathcal{H})$.
Then we get a natural composition rule
\begin{equation*}
 (f \star g) (x,y) = \sum_{n,l,m=1}^\infty | n \rangle \langle n|
 f (x,y) | l \rangle \langle l| g(x,y) | m \rangle \langle m|
      = \sum_{n,l,m=1}^\infty F_{nl}(x) G_{lm} (x) | n \rangle \langle m|.
\end{equation*}
The above composition rule implies that the ordering in the NC algebra $\mathcal{A}^d_\theta$
is perfectly compatible with the ordering in the matrix algebra $\mathcal{A}^d_N$.
Thus we can straightforwardly translate multiplications of NC fields in $\mathcal{A}^d_\theta$
into those of matrices in $\mathcal{A}^d_N$ using the matrix representation (\ref{matrix-rep})
without any ordering ambiguity. Using the map (\ref{matrix-rep}),
the trace over $\mathcal{A}^d_\theta$ can also be transformed into the trace over $\mathcal{A}^d_N$, i.e.,
\begin{equation}\label{matrix-trace}
    \int_M \frac{d^{2n} y}{(2 \pi )^n |\mathrm{Pf}\theta|} = \mathrm{Tr}_{\mathcal{H}} = \mathrm{Tr}_N.
\end{equation}

Using the matrix representation (\ref{matrix-rep}),
we can show \cite{hsy-jhep09,japan-matrix,nc-seiberg,hsy-epjc09} that the $D=(d+2n)$-dimensional
NC $U(1)$ gauge theory on $\mathbb{R}^{d-1,1} \times \mathbb{R}^{2n}_{\theta}$ is exactly mapped
to the $d$-dimensional $U(N \to \infty)$ Yang-Mills theory on $\mathbb{R}^{d-1,1}$:
\begin{eqnarray} \label{equiv-ncu1}
 S &=& - \frac{1}{G_{YM}^2} \int d^D Y \frac{1}{4}(\widehat{F}_{AB} - B_{AB})^2 \\
 \label{equiv-u1un}
   &=& - \frac{1}{g_{YM}^2} \int d^d x \mathrm{Tr} \Bigl( \frac{1}{4} F_{\mu\nu}F^{\mu\nu}
   + \frac{1}{2} D_\mu \phi_a D^\mu \phi_a - \frac{1}{4}[\phi_a, \phi_b]^2 \Bigr)
\end{eqnarray}
where $G_{YM}^2 = (2 \pi )^n |\mathrm{Pf}\theta| g_{YM}^2$ and
$$B_{AB} = \left(
                  \begin{array}{cc}
                    0 & 0 \\
                  0 & B_{ab} \\
                  \end{array}
                \right).$$
Note that the equivalence between the $D$-dimensional NC $U(1)$ gauge theory (\ref{equiv-ncu1})
and $d$-dimensional $U(N \to \infty)$ Yang-Mill theory (\ref{equiv-u1un}) is not a dimensional reduction
but an exact mathematical identity.
A remarkable point is that the large $N$ gauge theories described
by the action (\ref{equiv-u1un}) arise as a nonperturbative formulation of string/M theories \cite{wtaylor}.
For instance, we get the IKKT matrix model for $d=0$ \cite{ikkt}, the BFSS matrix quantum mechanics
for $d=1$ \cite{bfss} and the matrix string theory for $d=2$ \cite{dvv}.
The most interesting case arises for $d=4$ and $n=3$ which suggests an engrossing duality \cite{hea} that
the 10-dimensional NC $U(1)$ gauge theory on $\mathbb{R}^{3,1} \times \mathbb{R}^{6}_{\theta}$ is
equivalent to the bosonic action of 4-dimensional $\mathcal{N} = 4$ supersymmetric $U(N)$ Yang-Mills theory,
which is the large $N$ gauge theory of the AdS/CFT duality \cite{ads-cft1,ads-cft2,ads-cft3}.
According to the large $N$ duality or gauge-gravity duality, the resulting large $N$ gauge theory
must be dual to a higher dimensional gravity or string theory as summarized in Fig. \ref{fchart:emg}.
Hence it should not be surprising that the NC $U(1)$ gauge theory should describe a theory of
gravity (or a string theory) in the same dimensions. Unfortunately this important possibility
has been largely ignored until recently in spite of the apparent relationship depicted in Fig. \ref{fchart:emg}.

\begin{figure}
\centerline{\psfig{file=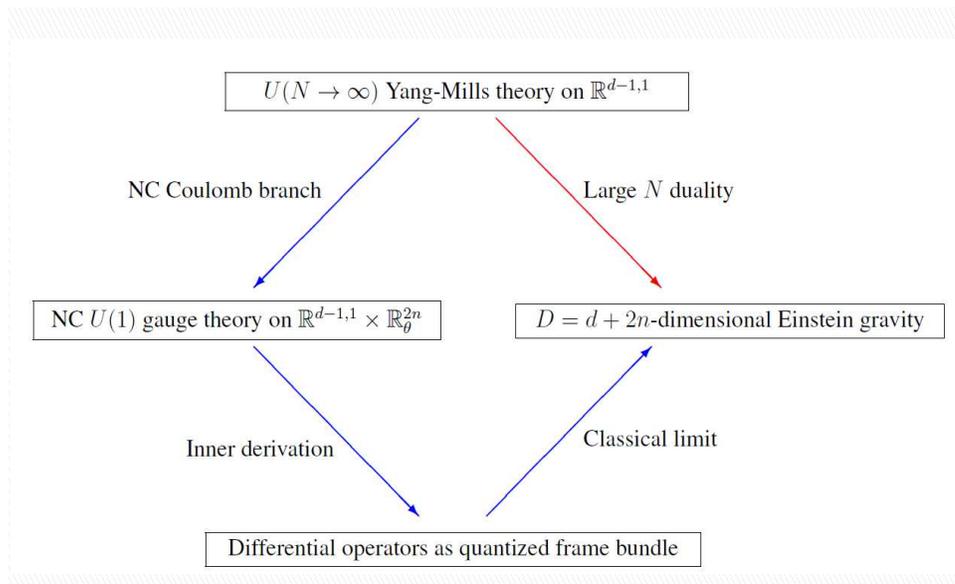,width=12.7cm}}
\vspace*{18pt}
\caption{Flowchart for large $N$ duality \cite{part1}}
\label{fchart:lnd}
\end{figure}

The matrix representation in Fig. \ref{fchart:emg} is based on the fact
that the NC space \eq{extra-nc2n} admits a separable Hilbert space and NC $U(1)$ gauge fields become
operators acting on the Hilbert space. Therefore there should be some way to map the NC $U(1)$ gauge
theory to the Einstein gravity if we accept the conjectural large $N$ duality.
The blue arrows in Fig. \ref{fchart:emg} show how to derive $D$-dimensional Einstein gravity
from NC $U(1)$ gauge theory on $\mathbb{R}^{d-1,1} \times \mathbb{R}_\theta^{2n}$,
which should be expected according to the (conjectural) large $N$ duality.
However we can use the emergent gravity from NC $U(1)$ gauge theory to verify the
large $N$ duality by realizing the equivalence between the actions (\ref{equiv-ncu1})
and (\ref{equiv-u1un}) in a reverse way. It is based on the observation \cite{q-emg,hea}
that the NC space \eq{extra-nc2n} is a consistent vacuum solution of the $d$-dimensional
$U(N \to \infty)$ Yang-Mills theory (\ref{equiv-u1un}) in the Coulomb branch.
See Fig.  \ref{fchart:lnd}.
The conventional choice of vacuum in the Coulomb branch of $U(N)$ Yang-Mills theory is given by
\begin{equation}\label{c-coulomb}
[\phi_a, \phi_b]|_{\mathrm{vac}} = 0 \qquad \Rightarrow \qquad
\langle \phi_a \rangle_{\mathrm{vac}} = \mathrm{diag}\big( (\alpha_a)_1, (\alpha_a)_2,
\cdots, (\alpha_a)_N \big)
\end{equation}
for $a=1, \cdots, 2n$. In this case the $U(N)$ gauge symmetry is broken to $U(1)^N$.
If we consider the $N \to \infty$ limit, the large $N$ limit opens a new phase of the Coulomb
branch given by
\begin{equation}\label{nc-coulomb}
    [\phi_a, \phi_b]|_{\mathrm{vac}} = - i B_{ab} \qquad \Rightarrow \qquad
    \langle \phi_a \rangle_{\mathrm{vac}} = p_a \equiv B_{ab} y^b
\end{equation}
where $B_{ab} = (\theta^{-1})_{ab}$ and the vacuum moduli $y^a$ satisfy the Moyal-Heisenberg
algebra \eq{extra-nc2n}. This vacuum will be called the NC Coulomb branch.
Note that the Moyal-Heisenberg vacuum \eq{nc-coulomb} saves the NC nature
of matrices while the conventional vacuum \eq{c-coulomb} dismisses the property.
Suppose that fluctuations around the vacuum \eq{nc-coulomb} take the form
\begin{equation}\label{gen-sol}
 D_\mu = \partial_\mu - i \widehat{A}_\mu (x, y), \qquad \phi_a = p_a + \widehat{A}_a (x, y).
\end{equation}
The adjoint scalar fields in Eq. \eq{gen-sol} now obey the deformed algebra given by
\begin{equation}\label{def-moyal}
    [\phi_a, \phi_b] = - i (B_{ab} - \widehat{F}_{ab}),
\end{equation}
where
\begin{equation}\label{field-st}
    \widehat{F}_{ab} = \partial_a \widehat{A}_b - \partial_b \widehat{A}_a
    - i [\widehat{A}_a, \widehat{A}_b]
\end{equation}
with the definition $\partial_a \equiv \mathrm{ad}_{p_a} = -i [p_a, \cdot]$.
Plugging the fluctuations in Eq. \eq{gen-sol} into the $d$-dimensional $U(N \to \infty)$
Yang-Mills theory (\ref{equiv-u1un}), we finally get the $D=(d+2n)$-dimensional
NC $U(1)$ gauge theory and thus arrive at the reversed version of
the equivalence \cite{hea,q-emg}:
\begin{eqnarray} \label{rev-equiv}
 S &=& - \frac{1}{g_{YM}^2} \int d^d x \mathrm{Tr} \Bigl( \frac{1}{4} F_{\mu\nu}F^{\mu\nu}
   + \frac{1}{2} D_\mu \phi_a D^\mu \phi_a - \frac{1}{4}[\phi_a, \phi_b]^2 \Bigr) \nonumber \\
&=& - \frac{1}{G_{YM}^2} \int d^D Y \frac{1}{4}(\widehat{F}_{AB} - B_{AB})^2,
\end{eqnarray}
where $\widehat{A}_A (x, y) = (\widehat{A}_\mu, \widehat{A}_a) (x, y)$ are $D=(d+2n)$-dimensional
NC $U(1)$ gauge fields. It might be emphasized that the NC space \eq{nc-coulomb} is
the crux to realize the equivalence \eq{rev-equiv}. If the conventional commutative vacuum \eq{c-coulomb}
were chosen, we would have failed to realize the equivalence \eq{rev-equiv}.
Indeed it turns out \cite{hea} that the NC Coulomb branch is crucial to realize the emergent gravity
from matrix models or large $N$ gauge theories as depicted in Fig. \ref{fchart:lnd}.

The relationship between a lower-dimensional large $N$ gauge theory and a higher-dimensional
NC $U(1)$ gauge theory in Figs. \ref{fchart:emg} and \ref{fchart:lnd} is an exact mathematical identity.
Therefore the next step is to find an isomorphic map from the NC $U(1)$ gauge theory
to the Einstein gravity in order to complete the large $N$ duality.
To be precise, consider the inverse metric in Einstein gravity given by
\begin{equation}\label{i-metric}
    \Big( \frac{\partial}{\partial s} \Big)^2 = E_A \otimes E_A
    = g^{MN} (X) \partial_M \otimes \partial_N,
\end{equation}
where $E_A = E_A^M (X) \partial_M$ are orthonormal frames on the tangent bundle $T\mathcal{M}$ of
a $D$-dimensional spacetime manifold $\mathcal{M}$. The large $N$ (or gauge-gravity) duality
in Figs. \ref{fchart:emg} and \ref{fchart:lnd} means to realize the vector
fields $E_A = E_A^M (X) \partial_M \in \Gamma (T\mathcal{M})$ in terms of NC $U(1)$ gauge fields.

It was shown in Ref. \refcite{part1} that quantum gravity can be derived from the electromagnetism
on NC spacetime by realizing the large $N$ duality in Fig. \ref{fchart:lnd} via the duality chain
given by
\begin{equation}\label{dd-chain}
  \mathcal{A}^d_N  \quad \Longrightarrow \quad \mathcal{A}^d_\theta \quad
  \Longrightarrow \quad \mathfrak{D}^d.
\end{equation}
See the paragraph in Eq. \eq{matrix-rep} for the definition of $\mathcal{A}^d_N$ and
$\mathcal{A}^d_\theta$. The module of derivations $\mathfrak{D}^d$ is a direct sum of
the submodules of horizontal and inner derivations \cite{azam}:
\begin{equation}\label{d-deriv}
  \mathfrak{D}^d = \mathrm{Hor}(\mathcal{A}^d_N) \oplus \mathfrak{D} (\mathcal{A}^d_N) \cong
  \mathrm{Hor}(\mathcal{A}^d_\theta) \oplus \mathfrak{D} (\mathcal{A}^d_\theta),
\end{equation}
where horizontal derivation is locally generated by a vector field
\begin{equation}\label{dhor-vec}
k^\mu (x, y) \frac{\partial}{\partial x^\mu} \in \mathrm{Hor}(\mathcal{A}^d_\theta).
\end{equation}
In particular we are interested in the derivation algebra generated by the dynamical variables
in Eq. \eq{gen-sol}. It is defined by
\begin{equation}\label{der-d}
    \widehat{V}_A = \{ i \, \mathrm{ad}_{D_A} = [D_A, \; \cdot \;] | D_A (x,y) = (D_\mu, D_a) (x,y) \}
    \in \mathfrak{D}^d
\end{equation}
where $D_A (x,y) = - i \phi_A (x,y)$.
It can be shown that the duality chain in Eq. \eq{dd-chain} is constructed by
the quantization of a (regular) $d$-contact manifold \cite{vaisman2}.
The dynamical variables in $d$-dimensional Yang-Mills gauge theory in Fig. \ref{fchart:lnd}
take values in $\mathcal{A}_N^d$ while those in $D=(d+2n)$-dimensional NC $U(1)$ gauge theory
take values in $\mathcal{A}_\theta^d$. These two NC algebras $\mathcal{A}_N^d$ and
$\mathcal{A}_\theta^d$ are related to each other by considering the NC Coulomb branch
for the algebra $\mathcal{A}_N^d$ \cite{part1}.

In a large-distance limit, i.e. $|\theta| \to 0$, one can expand the NC vector fields $\widehat{V}_A$
in Eq. \eq{der-d} using the explicit form of the Moyal $\star$-product. The result takes the form
\begin{equation}\label{polyvector}
  \widehat{V}_A = V^M_A (x, y) \frac{\partial}{\partial Y^M} + \sum_{p=2}^\infty
  V^{a_1 \cdots a_p}_A (x, y) \frac{\partial}{\partial y^{a_1}} \cdots
  \frac{\partial}{\partial y^{a_p}} \in \mathfrak{D}^d,
\end{equation}
where $V^\mu_A = \delta_A^\mu$.
Thus the Taylor expansion of NC vector fields in $\mathfrak{D}^d$ generates an infinite tower of
the so-called polyvector fields which describe higher-spin fields with spin $s \geq 2$ \cite{q-emg}.
Note that the leading term gives rise to
the ordinary vector fields that will be identified with a frame basis associated
with the tangent bundle $T\mathcal{M}$ of an emergent spacetime manifold $\mathcal{M}$.
It is important to perceive that the realization of emergent geometry through the derivation algebra
in Eq. \eq{der-d} is intrinsically local. Therefore it is necessary to consider patching or gluing
together the local constructions to form a set of global quantities. We will assume that local
coordinate patches have been consistently glued together to yield global (poly)vector fields.
See Refs. \refcite{nc-glue1,nc-glue2} for a global construction of NC $\star$-algebras and
Ref. \refcite{q-emg} for the globalization of emergent geometry.
Let us truncate the above polyvector fields to ordinary vector fields given by
\begin{equation}\label{lorentzian-vec}
 \mathfrak{X}(\mathcal{M}) = \Big\{ V_A = V_A^M (x, y) \frac{\partial}{\partial X^M}|
 A, M = 0, 1, \cdots, D-1 \Big\}
\end{equation}
where $X^M = (x^\mu, y^a)$ are local coordinates on a $D$-dimensional emergent {\it Lorentzian}
manifold $\mathcal{M}$. The orthonormal vielbeins on $T\mathcal{M}$ are then
defined by the relation
\begin{equation}\label{lorentz-iviel}
    V_A = \lambda E_A \in \Gamma(T\mathcal{M})
\end{equation}
or on $T^* \mathcal{M}$
\begin{equation}\label{lorentz-viel}
    v^A = \lambda^{-1} e^A \in \Gamma(T^* \mathcal{M}).
\end{equation}
The conformal factor $\lambda \in C^\infty (\mathcal{M})$ is determined by the volume
preserving condition
\begin{equation}\label{tvol-cond}
    \mathcal{L}_{V_A} \nu_t = \big( \nabla \cdot V_A + (2-d-2n) V_A \ln \lambda \big) \nu_t = 0,
    \qquad \forall A = 0, 1, \cdots, D-1,
\end{equation}
where
\begin{equation}\label{vol-t}
    \nu_t \equiv d^d x \wedge \nu = \lambda^2  d^d x \wedge v^1 \wedge \cdots \wedge v^{2n}
\end{equation}
is a $D$-dimensional volume form on $\mathcal{M}$.
In the end, the Lorentzian metric on a $D$-dimensional spacetime manifold $\mathcal{M}$
is given by \cite{hsy-jhep09,hsy-review,q-emg}
\begin{eqnarray}\label{eml-metric}
    ds^2 &=& \mathcal{G}_{MN} (X) dX^M \otimes dX^N = e^A \otimes e^A \nonumber \\
    &=& \lambda^2 v^A \otimes v^A = \lambda^2 \big(\eta_{\mu\nu} dx^\mu dx^\nu
    + v^a_b v^a_c (dy^b - \mathbf{A}^b)    (dy^c - \mathbf{A}^c) \big)
\end{eqnarray}
where $\mathbf{A}^b := A_\mu^b (x, y) dx^\mu$. Therefore the NC field theory representation of
the $d$-dimensional large $N$ gauge theory in the NC Coulomb branch provides a powerful machinery
to identify gravitational variables dual to large $N$ matrices.
Therefore we see that the general large $N$ duality depicted in Fig. \ref{fchart:lnd}
can be explained with the map \eq{dd-chain} \cite{hsy-jhep09,part1,q-emg}.

A crucial point is that the NC space \eq{extra-nc2n} is caused by the Planck energy condensate
in vacuum and responsible for the generation of flat spacetime.
The dynamical origin of flat spacetime in emergent gravity is a tangible difference
from general relativity and the crux to clearly resolve the notorious cosmological constant
problem \cite{hsy-review,hsy-jhep09,hsy-jpcs12}.
A remarkable feature is that the huge vacuum energy being a perplexing cosmological constant
in general relativity was simply used to generate flat spacetime and thus does not gravitate.
Since the Planck energy condensate into vacuum must be a dynamical process,
it is natural to expect that the cosmic inflation corresponds to a dynamical mechanism
for the instantaneous condensation of vacuum energy to enormously spread out spacetime.
Recently it was shown in Ref. \refcite{part1} that the cosmic inflation actually arises as
a time-dependent solution of the matrix quantum mechanics (i.e., $d=1$ case
in Eq. \eq{rev-equiv}) describing the dynamical process of the vacuum condensate
without introducing any inflaton field as well as an {\it ad hoc} inflation potential.
The emergent spacetime picture admits a background-independent
formulation so that the inflation is responsible for the dynamical emergence of spacetime.
These results require us to seriously reconsider main sources to introduce the multiverse hypothesis
from the standpoint of emergent spacetime.
The emergent spacetime picture certainly opens a new perspective that may cripple all the rationales
to introduce the multiverse hypothesis such as the cosmological constant problem,
chaotic and eternal inflation scenarios, and string landscape \cite{hsy-essay}.

Although string theory has a high potential for TOE, it has failed to achieve the goal so far.
We think that a main reason for the failure lies on the incapacity for generating space and time
through the vibrations of a string without assuming the prior existence of spacetime.
We argued that, from the deformation theory point of view,
replacing particles (point-like objects) with strings (one-dimensional objects) is equivalent
to doing commutative space by NC space.
Then the gravitational force manifests itself as the deformation of NC algebra via
NC U(1) gauge fields, leading to a dynamical NC spacetime and realizing emergent spacetime \cite{part1}.
The whole picture in Fig. \ref{fchart:lnd} thus implies that NC spactime can be viewed
as a second-quantized string and demonstrates how the deformation \eq{extra-nc2n} provides
us a background-independent formulation of string theory.

\section*{Acknowledgments}

We thank organizers of the 2nd LeCosPA Symposium, especially Pisin Chen for invitation and hospitality.
This work was supported by the National Research Foundation of Korea (NRF) grant
funded by the Korea government (MOE) (No. 2011-0010597 and No. NRF-2015R1D1A1A01059710).

\end{document}